\documentclass[twocolumn,showpacs,preprintnumbers,amsmath,amssymb]{revtex4}
\usepackage{txfonts}
\usepackage{graphicx}% Include figure files
\usepackage{dcolumn}% Align table columns on decimal point
\usepackage{bm}% bold math
\usepackage{verbatim}

\begin{document}
\title{ Traffic flow in a Manhattan-like urban system}

\author{Ming Li$^{a}$}
\author{Zhong-Jun Ding$^{a}$}
\author{Rui Jiang$^{b,c}$}\email{rjiang@ustc.edu.cn}
\author{Mao-Bin Hu$^{b}$}
\author{Bing-Hong Wang$^{a,d}$}\email{bhwang@ustc.edu.cn}

\affiliation{$^{a}$ Department of Modern Physics, University of
Science and Technology of China, Hefei, 230026, People's Republic
of China}

\affiliation{$^{b}$ School of Engineering Science, University of
Science and Technology of China, Hefei, 230026, People's Republic
of China}

\affiliation{$^{c}$ State Key laboratory of Fire Science,
University of Science and Technology of China, Hefei, 230026,
People's Republic of China}

\affiliation{$^{d}$ Complex System Research Center, University of
Shanghai for Science and Technology and Shanghai Academy of System
Science, Shanghai, 200093, People's Republic of China}

\date{\today}

\begin{abstract}
In this paper, a cellular automaton model of vehicular traffic in
Manhattan-like urban system is proposed. In this model, the
origin-destination trips and traffic lights have been considered.
The system exhibits three different states, i.e., moving state,
saturation state and global deadlock state. With a grid coarsening
method, vehicle distribution in the moving state and the
saturation state has been studied. Interesting structures (e.g.,
windmill-like one, T-shirt-like one, Y-like one) have been
revealed. A metastability of the system is observed in the
transition from saturation state to global deadlock state. The
effect of advanced traveller information system (ATIS), the
traffic light period, and the traffic light
switch strategy have also been investigated. %It is found that with
%the help of ATIS and by choosing proper traffic light period and
%strategy, the system could become more stable (i.e., the
%transition to global deadlock state could happen at larger
%density).

%The system exhibits a sharp transition from moving state to
%jamming state. By using mean velocity feedback routing strategy,
%the urban traffic capacity (measured by phase transition point)
%increases significantly and the moving state is expanded to a
%second phase. With a grid coarsening method, vehicle distribution
%is studied and compared. A non-monotonic variation of urban
%traffic capacity with the mean velocity search length is captured.
%The dependence of mean velocity and critical density on traffic
%light period are also studied.
\end{abstract}


\maketitle

\section{INTRODUCTION}

In modern society, the transportation of people and goods as well
as information are becoming more and more frequent. As a result,
in transportation and communication systems, traffic congestion
has become one of the urgent issues to
be tackled [1-3]. %Moreover, traffic flow exhibits various nonlinear and
%nonequilibrium phenomena, e.g., boundary induced phase transition,
%metastable state and hysteresis, among others. Therefore, it has
%attracted the interests of the physics community.
Recent researches in the field of network traffic of  information
packet indicate that the network capacity could be remarkably
improved by optimizing the routing strategy [4-10]. Therefore, it
is natural to expect that the efficiency of transportation network
of vehicles could also be improved in the same way, in particular
with the help of advanced traveller information system (ATIS).

Recently, Scellato et al have studied the vehicular flow in urban
street network \cite{routing}. They have investigated a congestion
aware routing strategy, which allows the vehicles to dynamically
update the routes towards their destinations. It is shown that in
real urban street network of various cities, a global traffic
optimization could be achieved based on local agent decisions.

Nevertheless, we would like to point out that in the work of
Scellato et al., the influence of traffic lights has not been taken
into account. In their model, the vehicles coming from different
adjacent streets compete for the same intersection. Since traffic
lights play a very important role in urban traffic, their effect
needs to be carefully investigated. Moreover, by using the routing
strategy in Ref.\cite{routing}, another deficiency is that there
might be some vehicles which could never reach their destinations
(or it takes extremely long time to reach the destinations) because
they are always detouring around the destinations.

%proposed %a quite realistic model to
%study network traffic, which is out of the framework of BML model.

%both local and
%global routing strategies~\cite{routing}. %  were studied to improve traffic conditions
%But they did not consider traffic lights, so that vehicles coming
%from different adjacent streets compete for the same intersection.
%This is not in consistence with reality, and the impacts of
%traffic lights can not be reflected.
% For urban traffic system, the structural elements of city networks exert an immense influence onto the traffic dynamics.

In this paper, we propose a model to study urban traffic
considering the traffic lights. For simplicity, our model is
implemented in a Manhattan-like urban system. In such system,
there are usually more than one shortest path between an origin
and destination pair \footnote{Sometimes there is only one
shortest path, e.g., the origin and the destination are on same
row or column. }. Therefore, also for simplicity, vehicles are
only allowed to go along shortest paths in this paper. The routing
strategy determines which shortest path to choose at the
intersections. %Also for simplicity reason, this paper only
%investigates the synchronous traffic lights case.
It is shown that three different states are observed in our model.
%a sharp transition to global deadlock occurs in our model.
With the help of ATIS and by choosing proper traffic light period,
the transition to global deadlock state could happen at larger
density.

%The traffic rules are based on BML model and NS model. We study
%the effects of mean velocity feedback routing strategy. By using
%mean velocity feedback routing strategy, the urban traffic
%capacity is greatly improved. With a grid coarsening method,
%vehicle distribution is show to be consistent with reality. The
%effects of mean velocity search length and traffic light period
%are also studied.

This paper is organized as follow. The urban traffic model and
routing strategy will be presented in section II. In section III,
the simulation results will be discussed. Finally, we present the
conclusion and outlook in section IV.

\section{MODEL}

In our model, urban city is presented as a Manhattan-like system
as shown in Fig.1(a). We model the network of streets as a $N
\times
N$ square lattice. %It is assumed that if a vehicle occupies an
%intersection, other vehicles can not go through the intersection.
The spatial separation between any two successive intersections is
assumed to have one lane for each direction, and each lane is
divided into $L$ cells. Vehicle driving is restricted to the right
lane. % (or left lane in Japan, U.K. and some other countries). % between two successive intersections.
%In order to avoid several vehicles coming from adjacent streets
%compete for the same intersection,
On each lane, the cells are numbered $1,2,\ldots,L$ from
downstream to upstream. For simplicity, we assume that synchronous
traffic lights with fixed period are set at each intersection. In
each traffic phase, the traffic lights keep green for one ingoing
street and red for other three ingoing streets for $T$ time steps.
Therefore, the traffic light period is $4T$ (the yellow light is
not considered). When the green light is on, vehicles on the
corresponding ingoing street could go straight ahead, turn left or
right, or make a U-turn. In this way, the conflicts between
turning vehicles and straight vehicles are avoided. Although our
setup is simplified, it could capture the essence of traffic
congestion in urban area, i.e., the spillover of queues from
downstream intersection to upstream ones.
%The traffic lights change synchronously as follows
%\footnote{The movie in the supplement material shows details of
%evolution of traffic lights and movement of vehicles in a small
%scale system.}. All traffic lights keep green for a same direction
%and red for the other three directions for $T$ steps. The green
%light changes simultaneously in a clockwise manner at each
%intersection. Therefore, the traffic light period is $4T$.

\begin{figure}
\scalebox{0.4}[0.4]{\includegraphics{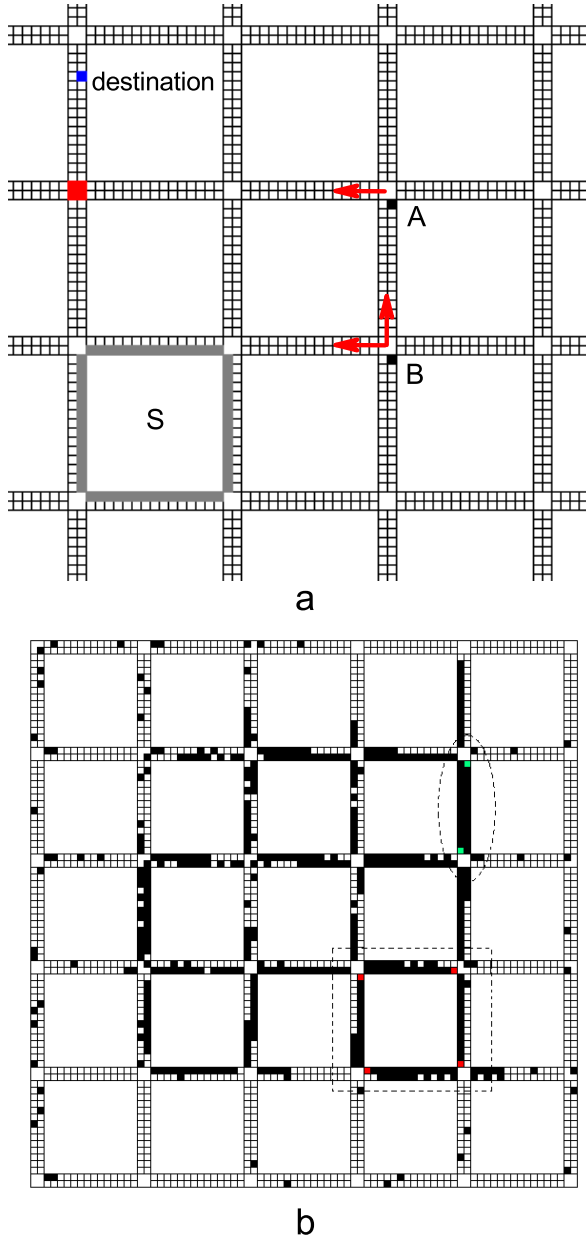}} \caption{(Color
online) (a) Example of a Manhattan-like urban system. Each cell
can either be empty or occupied by a vehicle. The black sites are
occupied by vehicle A and B. The blue site is their destination.
As right-hand side driving, the vehicles must go to the red
intersection first. Currently, there are two directions for
vehicle B to choose, but only one direction for vehicle A (shown
by red arrows). The vehicle density of lattice $S$, is the average
vehicle density of the four roads (grey roads) around $S$. (b)
Snapshot of two local deadlocks in a small-sized system. The loop
structure of the local deadlock is indicated by the dashed lines.
The four leading vehicles (indicated by red) in the square loop
wish to make right turn; the two leading vehicles (indicated by
green) in the oval loop wish to make U-turn.  }
\end{figure}

The state of the system is updated % at time $t+1$ can be obtained from that at time $t$
by applying the Nagel-Schreckenberg (NS) rules \cite{ns} to each
vehicle in parallel, except the leading vehicles on each road. The
NS rules are as follow:
\begin{itemize}
\item Acceleration: $v_n \rightarrow  \min(v_n + 1,v_{max})$;

\item Deceleration: $v_n \rightarrow \min(v_n,d_n)$;

\item Random brake: $v_n \rightarrow \max(v_n-1,0)$ with a braking
probability $p$;

\item Movement: $x_n \rightarrow x_n+v_n $;

\end{itemize}
where $v_{max}$ is the maximum velocity of vehicles, $x_n$ is the
position of the $n$th vehicle on each road and $d_n=x_{n+1}-x_n-1$
is the distance to the vehicle ahead.

In our model, once an outgoing street is in jam, the vehicles
choosing that outgoing street are not allowed to enter the
intersection to avoid hindering vehicles on other streets. We
assume that an outgoing street is in jam once its last two cells
are simultaneously occupied.

As a result, update rules of the leading vehicles are the same as
NS rules except the definition of $d$. The details are listed below: \\
$(1)$ Traffic light is green
\begin{itemize}
  \item if the desired outgoing street is in jam or the intersection is occupied by other vehicle which
  needs additional time to drive into its desired outgoing street (if the yellow traffic light period is taken into
   account, the latter situation would hardly occur),
  $d$ is the distance to the intersection;
  \item otherwise, $d$ is the distance to the last vehicle of the desired outgoing street.
\end{itemize}
$(2)$ Traffic light is red
\begin{itemize}
  \item  $d$ is the distance to the intersection.
\end{itemize}

Each vehicle has its origin and destination. Initially, all
vehicles randomly select their origin and destination. When a
vehicle reaches its destination, a new destination is chosen
randomly from the system beyond the current road. Thus the number
of vehicles is conserved. As mentioned before, in our model
vehicles are only allowed to go along shortest paths, but the
drivers need to determine which outgoing street to run into at
each intersection when it is necessary. As shown in Fig.1(a), when
a vehicle reaches an intersection, two situations could appear:
the vehicle either has two directions (vehicle B) or only one
direction (vehicle A) to choose. %The former situation happens even
%more frequently than the latter one.
We assume that without the
help of ATIS, the driver will choose a direction randomly in the
former situation.

Nowadays, ATIS could provide real-time information to the drivers.
We suppose that at each intersection, the ATIS provides mean
velocity %within last $l$ cells
on each outgoing street, and the
drivers will choose the direction with larger
value of mean velocity \cite{mvfs}. %When the two outgoing streets have equal value of
%$v$, the drivers will choose one randomly.
%We would like to mention that in Eq.(1) in Ref.[?], the density
%information (fraction of occupied cells) is used in routing
%strategy. We assume that ATIS provides mean velocity information
%in our model because the expected travel time could be estimated more conveniently via mean velocity. %for an ATIS, it is much
%easier to collect mean velocity information than density
%information in real traffic.
We have also tested the usage of density information and
congestion coefficient information as in Refs.\cite{routing,
ccfs}, and similar results could be obtained.

\section{SIMULATION AND DISCUSSION}

All the  simulation results shown here are obtained after
discarding the first $10^5$ time steps (as transient time) and
then averaging over the next $10^4$ time steps. The system size is
$N\times N=24\times 24$, and the length of street is $L=100$. In
NS rules, the maximum velocity of vehicles is $v_{max}=3$ and the
probability of random braking is $p=0.1$. %We first consider the
%special case of search length $l=L$ and traffic light period
%$4T=80$. The effects of varying $l$ and $T$ will be discussed
%later.
The traffic light
period parameter $T$ is set to 20, unless otherwise mentioned.

%In order to demonstrate the traffic characteristics of the system,
%the fundamental diagrams of the system are shown in

%\subsection{Synchronous traffic lights}

\subsection{Clockwise strategy}

This subsection studies the situation that the green lights change
simultaneously in a clockwise manner at all intersections. Fig.2
shows the average velocity and the average flux of each run in
system without ATIS. The average flux $\langle f\rangle$ of the
system is defined as the average flux of all roads in the system.
One can see that three different states could be identified.

\begin{figure}
\scalebox{0.4}[0.4]{\includegraphics{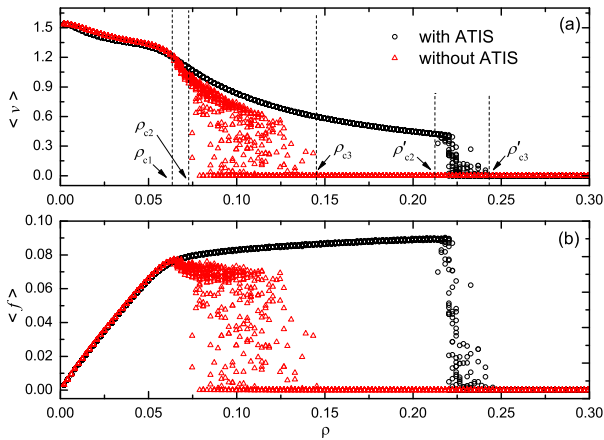}} \caption{(Color
online) The average velocity (a) and the average flux (b) of each
run in the system without and with ATIS. }
\end{figure}

\begin{figure*}
\scalebox{0.8}[0.8]{\includegraphics{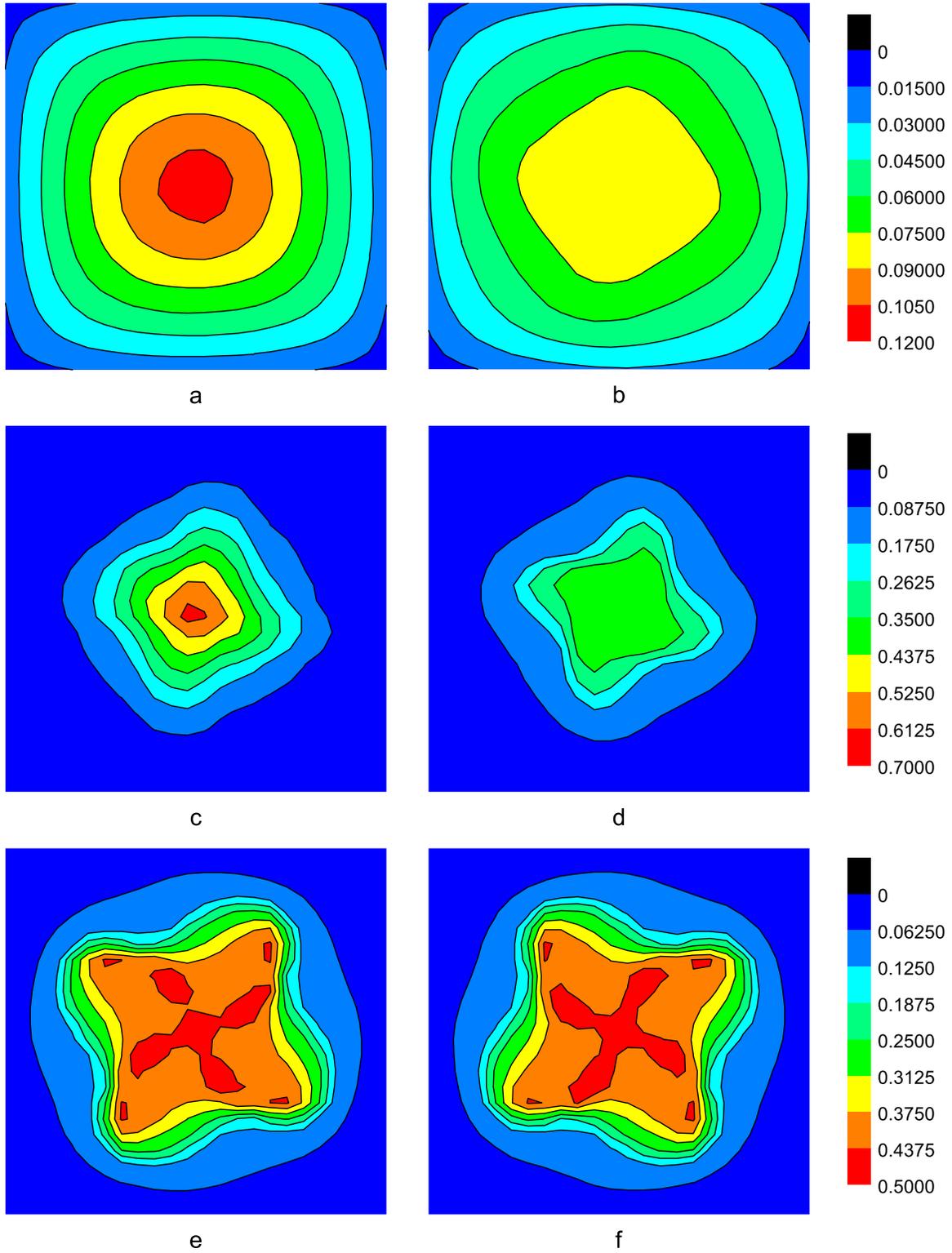}} \caption{(Color
online) Typical patterns of the distribution of vehicles. (a), (b)
$\rho=0.05$; (c), (d) $\rho=0.1$; (e), (f) $\rho=0.15$. In (a) and
(c), system has no ATIS; In (b) and (d-f), system has ATIS. In
(a-e), the traffic lights are switched in clockwise manner; in (f),
the traffic lights are switched in anti-clockwise manner. Note the
scale of the colors is different at different rows.}
\end{figure*}

When the density is smaller than a critical density
$\rho_{c1}\approx0.063$, the system is in moving state, in which
the average flux  almost linearly increases with the increase of
density. Fig.3(a) displays typical pattern of the distribution of
vehicles in the moving state, employing a grid coarsening method
in which vehicle density of lattice S is defined as the average
vehicle density of the four roads around the lattice (see
Fig.1(a)). It can be seen that distribution of vehicles is
heterogeneous, more vehicles accumulate in the center. The inner
rings are
circles. The outer rings gradually become squares. %With the
%increase of system density $\rho$, the local density in the inner
%rings increases accordingly.

When the system density is in the range
$\rho_{c1}<\rho<\rho_{c2}\approx 0.073$, the system is in a
saturation state, in which the flow rate almost saturates (it only
slightly changes with the increase of density). When the system
density exceeds a third critical density $\rho_{c3}\approx 0.145$,
the local density in the center area becomes so large that a local
deadlock will be induced. {  The local deadlock exhibits loop
structure (see Fig.1(b)) in which all the desired outgoing streets
of leading vehicles on the ingoing streets are in jam. } This
deadlock spreads to the system and causes global deadlock, in
which no vehicle can move.

When the density is in the intermediate range
$\rho_{c2}<\rho<\rho_{c3}$, the system is metastable: it could
either evolve into global deadlock state or be in a saturation
state. The scattered data indicate that the system is transiting
from saturation state to global deadlock state in the data
collection period (from time step $10^5$ to $1.1\times10^5$).
Fig.3(c) shows typical pattern of the distribution of vehicles in
the saturation state at $\rho=0.1$. One can see that the local
density $\rho_l$ in the center area could reach very large value
($\rho_l\approx 0.7$). Moreover, the shape of the outer rings
gradually changes
from squares into diamonds. % and the whole pattern is gradually
%transiting into a four-angle-star structure. However, before the
%structure becomes more manifest, global deadlock will be induced.
%In contrast, in the system with ATIS, the four-angle-star
%structure can be observed clearly as shown below.

Now we study the effect of ATIS. Fig.2 also shows the average
velocity and the average flux of each run in system with ATIS.
Similarly, the three states mentioned above are observed.
Nevertheless, the system becomes much more stable with the help of
ATIS: The saturation state is stable in the density range
$\rho_{c1}<\rho<\rho_{c2}'\approx 0.212$. Moreover, the average
velocity and the flow rate of the saturation state are also
enhanced with the help of ATIS. The global deadlock state occurs
when the density $\rho>\rho_{c3}'\approx 0.243$. The system is
metastable in the density range $\rho_{c2}'<\rho<\rho_{c3}'$,
which is much narrower than that in system without ATIS.

Fig.3(b) shows the distribution of vehicles in the moving state in
the system with ATIS. Since the ATIS could help drivers to avoid
center area, the accumulation of vehicles in the center is
suppressed: the local density in the center area becomes much
smaller (c.f.Fig.3(a)). Moreover, different from the system
without ATIS, the inner ring exhibits diamond instead of circle.

Fig.3(d) shows the distribution of vehicles in the saturation
state at $\rho=0.1$. A four-angle-star structure begins to appear.
More interestingly, with the further increase of density, a
windmill-like high density structure appears (Fig.3(e)). The
structure exhibits remarkable deviation from the diagonal lines.
Due to symmetry, if the green lights change simultaneously in an
anti-clockwise manner, the structure will deviate from
the diagonal lines in an opposite way (see Fig.3(f)). %Figs.6(g) and (h) show the patterns, with the green light
%changes simultaneously in an anti-clockwise manner. One can see
%that a rotation of the star structure occurs, and it deviates from
%the diagonal lines in a clockwise way.
Appearance of the structure might be related to the betweenness
distribution because betweenness reaches the maximum on the
diagonal lines. However, more efforts are needed in future work to
explore the exact origin of the four-angle-star structure and the
windmill-like structure.

Next we investigate the influence of traffic light period. Fig.4
compares the average velocity of the system at different values of
$T$, in which the curves are obtained by averaging over many runs.
At small densities, the average velocity decreases with the
increase of $T$, with oscillations appearing (see Fig.5). The
oscillations could be analyzed by considering a minimal network
with one single intersection, as explained in details in
Ref.\cite{light}. Specifically, when traffic light period
parameter $T>T_f=L/(v_{\max}-p)=34.5$, the local minimum appears
at $T\approx nT_f$ ($n=1, 2, \cdots$). When $T<T_f$, the local
maximum appears at $T\approx T_f/n$ ($n=2, 3, \cdots$).

We also need to point out in the moving state, the average
velocity in the system with ATIS is smaller than that without
ATIS. The difference is very remarkable when $T$ is large (Fig.5),
but it gradually shrinks with the decrease of $T$. The difference
is due to the different distributions of vehicles (c.f.Figs.3(a)
and (b)). While ATIS decreases the value of local density in the
center area, the local density in the outer rings increases. While
the former effect tends to increase average velocity, the latter
one tends to decrease average velocity. In our case (Fig.5), the
latter effect prevails over the former one. Thus, the average
velocity decreases in system with ATIS.

Finally, we focus on the transition regime from saturation state
to global deadlock state. It can be seen that the transition
regime heavily depends on $T$ in both systems, and the transition
always occurs later with the help of ATIS. Moreover, the
transition is more sensitive to $T$ in the system with ATIS: In
the range $10\le T\le 40$, there exist three extremes (The
transition occurs later at the extremum than in the vicinity
range) at $T\approx 10, 20, 40$ in the system with ATIS while
there exists only one extremum at $T\approx 25$ in the system
without ATIS. We point out that the average velocity also reaches
maximum at $T\approx 10, 20, 40$ in the moving state (Fig.5).
Whether there exists an underlying relationship for the
coincidence or not needs to be explored in future work.

%It can be seen that in the system with ATIS, the transition regime
%reaches the

%With the increase of $T$, the transition regime from saturation
%state to global deadlock state changes non-monotonically in system
%both with and without ATIS. However, the two systems exhibit
%different dependence on $T$. While the transition regime exhibits
%several oscillations in the system with ATIS in the range $10\le
%T\le 40$, , it shows only once in the system without ATIS.

%Moreover, Fig.4 also shows that the average velocity in the moving
%phase {\it heavily} depends on $T$ in both systems, but it is
%essentially independent of $T$ in the saturation phase. Fig.5
%shows the plots of average velocity versus $T$ at density
%$\rho=?$, a typical value in moving phase. The average velocity
%decreases with the increase of $T$,  with oscillations with
%interval approximately equaling to $L/(v_{\max}-p)=34.5$, which
%could be analyzed by considering a minimal network with one single
%intersection, as explained in details in Ref.[?]. As for the
%oscillations happening at $T<30$, {\bf (why it happens, more text
%needed ......)}

\begin{figure}
\scalebox{0.4}[0.4]{\includegraphics{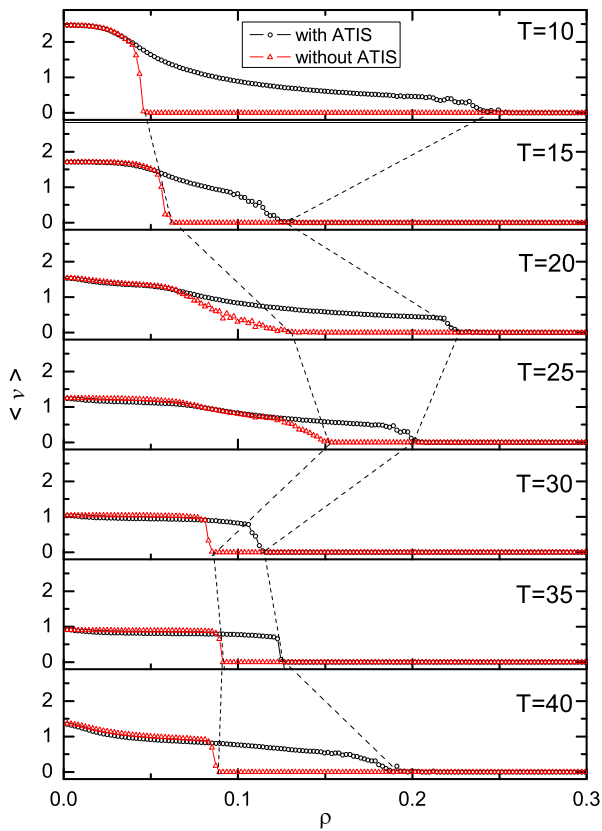}} \caption{(Color
online) The average velocity of the system at different values of
$T$. The dashed lines are guide for eyes, which roughly connect the
transition regime from saturation state to global deadlock state. }
\end{figure}

\begin{figure}
\scalebox{0.4}[0.4]{\includegraphics{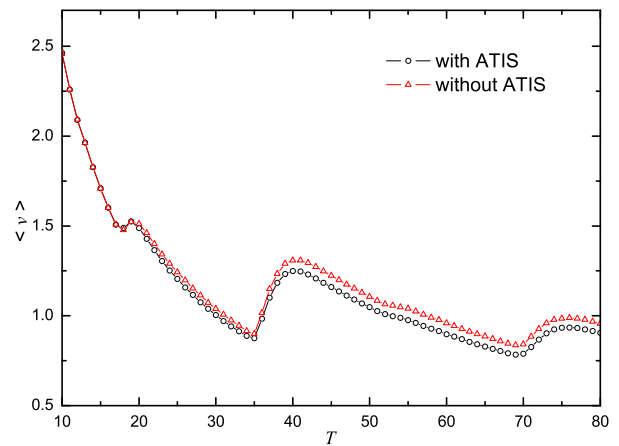}} \caption{(Color
online) The average velocity of the system at different values of
$T$. The system density $\rho=0.01$. }
\end{figure}

\subsection{8-like strategy}

In this subsection, we briefly study a different synchronous
traffic light strategy, in which the green lights are given
alternatively for the horizontal streets and the vertical streets.
Denote the traffic lights as 1, 2, 3, 4 in the clockwise manner,
the lights are switched in the order 1-3-2-4-1 or 1-3-4-2-1 in
this strategy. The former is named 8-like strategy and the latter
is named reverse 8-like strategy, since the switching way looks
like a ``8".

Our simulation shows that similar results could be obtained as in
previous subsection: the moving state, the saturation state and
the global deadlock could be observed. { Nevertheless, in the
system with ATIS}, the distribution of vehicles is much different.
Fig.6 shows that in the moving state and the saturation state, the
symmetry has been broken under the 8-like strategy. Instead of
four-angle-star structure and windmill-like structure, the
distribution exhibits Y-like structure (density between 0.525 and
0.6), T-shirt-like structure (density between 0.45 and 0.525)  and
triangle structure from inner to outer ring. { We also point out
that if the traffic lights are switched in the order 3-1-2-4-3 or
3-1-4-2-3, the vehicle distribution will exhibit symmetry with
respect to that shown in Fig.6. }

{On the other hand, in the system without ATIS, the distribution
of vehicles still remains symmetric as that under clockwise
strategy (not shown here). Therefore, although the exact formation
mechanism of various structures of the vehicle distribution is
still unclear, it at least depends on (i) traffic light switching
strategy, (ii) whether the ATIS is provided or not, and (iii) the
topology of the network. }
%The value of density in the inner most structure (i.e.,
%windmill-like structure and Y-like structure) affects the
%transition regime from saturation state to global deadlock. The
%one with larger value of density leads to earlier transition to
%global deadlock. Since the Y-like structure is smaller than
%windmill-like structure, it is

The traffic light switching strategy also has nontrivial impact on
the transition from saturation state to global deadlock. Fig.7
compares the average velocity of the 8-like strategy with that of
the clockwise strategy at four typical values of $T$. In system
with ATIS, 8-like strategy performs worse than clockwise strategy
except at $T=30$. { On the other hand, in system without ATIS,
8-like strategy never performs worse than clockwise strategy.}

\begin{figure*}
\scalebox{0.8}[0.8]{\includegraphics{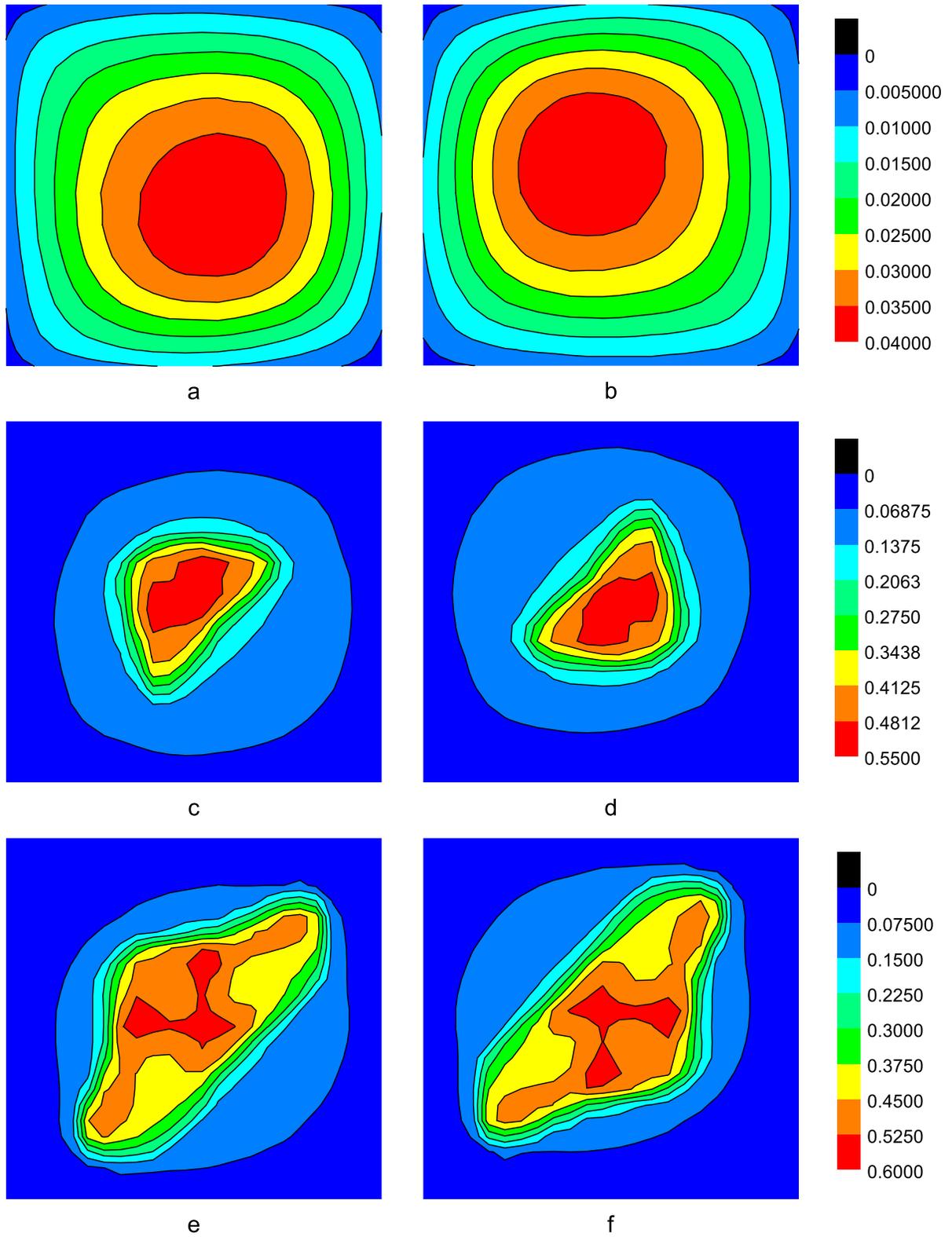}} \caption{(Color
online) Typical patterns of the distribution of vehicles in system
with ATIS. (a), (b) $\rho=0.02$; (c), (d) $\rho=0.1$; (e), (f)
$\rho=0.15$. In (a,c,e), 8-like strategy is adopted; in (b,d,f),
reverse 8-like strategy is adopted. Note the scale of the colors is
different at different rows.}
\end{figure*}

\begin{figure*}
\scalebox{0.5}[0.5]{\includegraphics{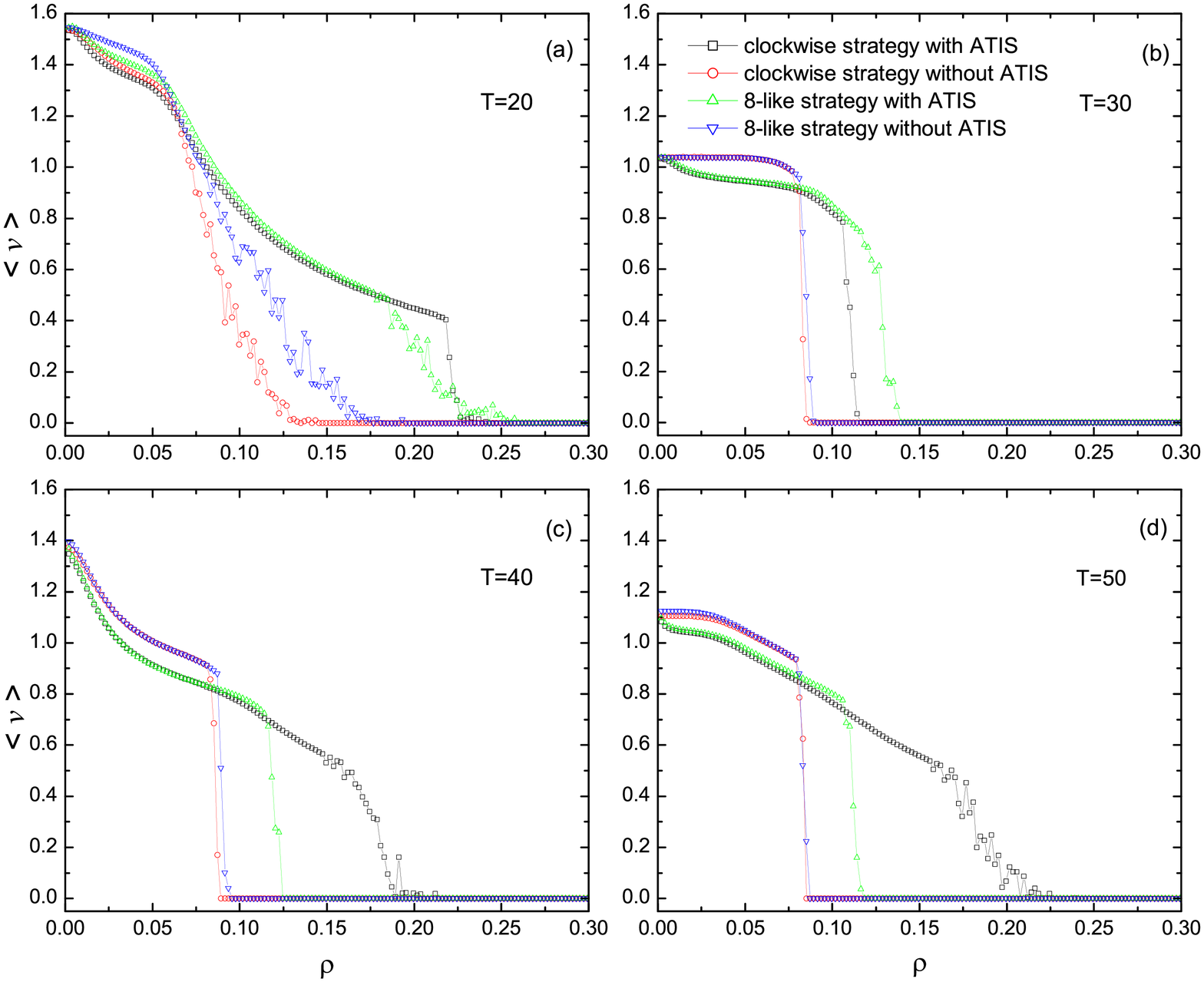}} \caption{(Color
online) Comparison of the average velocity of two different
strategies at four typical values of $T$.  }
\end{figure*}

%\subsection{Asynchronous Traffic Lights}

%This subsection studies asynchronous traffic lights. For
%simplicity, we assume that the traffic light periods are same at
%all intersection, and green light changes in a clockwise manner.
%Nevertheless, the switch time is randomly determined at each
%intersection.

%Our simulations show that similar results are observed as in the
%previous subsection. {\bf (more simulation results needed ......)}

\section{CONCLUSIONS}

In summary, a cellular automata model is proposed to study the
traffic flow in Manhattan-like urban system. We have considered
the origin-destination trips and traffic lights in this model. It
is found that the system could exhibit three different states,
i.e., the moving state in which the flow rate increases with the
density almost linearly, the saturation state in which the flow
rate only slightly changes with density, and the global deadlock
state in which no vehicle can move. With a grid coarsening method,
we explore the distribution of vehicles in the moving state and
the saturation state, which shows qualitatively different
structure. A metastability of the system is observed in the
transition from saturation state to global deadlock state. The
influence of ATIS, the traffic light period, and the traffic light
switch strategy on the system have been investigated.

%The structural
%elements of urban traffic, origin-destination trips and traffic
%lights are considered in this model. The system exhibits a first
%order transition from moving state to jamming state. When adopting
%the mean velocity feedback routing strategy, the system efficiency
%is greatly improved that the moving state is expanded to a much
%higher vehicle density. With information feedback, two phases
%emerge in moving state.
% We compare the two phases by simulation, and point out the
% differences of the vehicles distribution in the two phases.
%With a grid coarsening method, we find that the vehicle density
%shows a nonuniform distribution in the system, which is more
%closed to real situation. By using real time information feedback,
%vehicles are distributed more homogeneously, so the system
%efficiency is remarkably improved. Finally, the effects of mean
%velocity search length $l$ and traffic light period are also
%studied.

%relationship between critical density and search length $l$. The
%simulation results indicate that the driver need not to know the
%real-time information of the whole road to choose a better way. At
%last, for the sake of realizing the influence of traffic light,
%the fluctuation of average velocity with traffic light period is
%studied.

The simple model in this paper needs to be extended in several
directions in future work. (i) The routing strategy: optimal
strategy needs to be designed to further enhance the system
stability as well as the flow rate in the saturation state; (ii)
Traffic light strategy: other strategies (e.g., green wave
strategy, adaptive traffic light strategy) need to be investigated
or designed \cite{20,21,22}; (iii) System structure: more
realistic road network and multilane road sections need to be
considered; (iv) More realistic origin-destination data need to be
collected.

% We believe that it presents a general feature of urban traffic.
% Using this model, we investigate the dynamics of urban traffic,
% especially the effects of routing strategy.

%Although the routing strategy discussed in this paper utilizes
%local information, we can foresee that urban traffic efficiency
%will also be improved by using global information. In contrast to
%the global information, local information is easier to gather and
%calculate. Therefore, this strategy might be more practical in
%real-world applications. %This strategy might be applied in practice to improve urban traffic.
%Moreover, the vehicles are restricted to go along the shortest
%paths to their destinations in this paper. As we know, drivers
%often make detours when the nearby area is congested. If a
%suitable detour strategy is given, new phenomena may appear.
% which are worth of further study.

\section*{Acknowledgments}

This work is funded by the National Important Research
Project(Grant No.91024026); the National Natural Science
Foundation of China (Grant No.10975126, 11072239, 71171185), the
Specialized Research Fund for the Doctoral Program of Higher
Education of China (Grant No.20093402110032). R.J.acknowledges the
support of Fundamental Research Funds for the Central
Universities.

\end{document}